\documentclass[twocolumn,preprintnumbers,amsmath,amssymb,superscriptaddress]{revtex4}

\usepackage{amsmath}
\usepackage{amssymb}
\usepackage{bm}
\usepackage{booktabs}
\usepackage{graphicx}
\usepackage{multirow}
\usepackage{listings}
\usepackage{xcolor}
\definecolor{myblueold}{RGB}{65,105,225}
\definecolor{myblue}{RGB}{0, 71, 171}
\definecolor{mahogany}{RGB}{156, 0, 0}
\definecolor{mygray}{RGB}{145,145,145}
\definecolor{mybrown}{RGB}{110, 38, 14}
\definecolor{mygreen}{RGB}{0,128,0}
\newcommand{\bs}{\begin{subequations}}
\newcommand{\es}{\end{subequations}}
\usepackage[colorlinks=true, linkcolor=blue, citecolor=mygreen, urlcolor=blue]{hyperref}
\usepackage{textcomp}
\allowdisplaybreaks[1]

\newcommand{\ali}[1]{\begin{align}#1\end{align}}

\newcounter{RSQ}

\newcounter{DFQ}

\def\sh{\textcolor{myblue}{(sh)}}

\def\p{\textcolor{myblue}{(p)}}
\def\s{\textcolor{myblue}{(s)}}

\def\sc{\textcolor{myblue}{(sc)}}

\def\hvar{\textcolor{myblue}{h}}
\def\shvar{\textcolor{myblue}{sh}}
\def\svar{\textcolor{myblue}{s}}

\def\scvar{\textcolor{myblue}{sc}}

\newcommand{\ar}[1]{\textcolor{mybrown}{(#1)}}

\begin{document}

\setlength{\parskip}{0pt}

\title{\boldmath 
 QED corrections to bound-muon decays from an effective-field-theory framework}

\author{Duarte Fontes}
\affiliation{Institute for Theoretical Physics,
Karlsruhe Institute of Technology,
76128 Karlsruhe, Germany
}
\author{Robert Szafron}
\affiliation{Brookhaven National Laboratory, Upton, NY, U.S.A.
}

\date{\today}

\begin{abstract}
\noindent
Bound-muon decays are a powerful probe of new physics, making precise theoretical predictions for their spectra essential. While QED corrections significantly affect the shape of the spectra, their calculation is extremely challenging below the nuclear scale. By exploring the universality of modern effective-field-theory techniques, we present a framework that systematically computes those corrections across a broad class of bound-muon decays. As a key application, we provide the most accurate predictions to date for the signal and background spectra in muon conversion. We show that radiative corrections modify the leading-order ratio of these spectra by $5\%$ with minimal energy dependence, a result relevant for enhancing the discovery reach of upcoming experiments. Our framework also represents a crucial step toward connecting high-energy physics to low-energy observables, complementing recent progress above the muon mass scale.
\end{abstract}

\begin{flushright}
\vspace*{-9.5mm}
KA-TP-15-2025 \\
P3H-25-043
\end{flushright}
\vspace*{-0.1cm}

\maketitle
%


\noindent \textbf{Introduction.}
Muons serve as powerful probes of fundamental physics, enabling some of the most precise tests of the Standard Model (SM), such as measurements of the anomalous magnetic moment \cite{Athron:2025ets}. A particularly promising arena is that of bound muons (i.e. muons captured in atomic orbitals around nuclei), as they enable highly sensitive searches for charged lepton flavor violation (CLFV) \cite{Davidson:2022jai}. Two key processes in this context are muon conversion and muon decay-in-orbit (DIO). The former, which involves CLFV and has therefore not yet been observed, occurs when a muon converts into an electron in the field of a nucleus, producing a monoenergetic electron with energy near the muon mass. DIO refers to the SM process in which a bound muon decays into an electron and two neutrinos, yielding a continuous energy spectrum that extends up to energies overlapping with those expected from conversion electrons.

The exceptional appeal of these processes lies in their simplicity: the signal of muon conversion is sharp and well-defined, and its only irreducible background is DIO. This clean experimental signature is the basis for the ambitious goals of upcoming experiments such as Mu2e and COMET, which aim to improve sensitivity to muon conversion rates by four orders of magnitude 
\cite{Mu2e:2014fns, Diociaiuti:2024stz,COMET:2018auw, Fujii:2023vgo,Mu2e-II:2022blh}.
Yet this cleanliness is deceptive. In reality, QED corrections significantly reshape the spectra of both the signal and the background, altering the distributions in subtle but crucial ways. A precise understanding of these effects is essential for interpreting future results and fully exploiting the discovery potential of bound-muon experiments.

However, such effects are challenging to compute due to the presence of bound states. Unlike standard processes involving free particles, bound muons require a nonperturbative treatment of the Coulomb potential, placing the problem beyond the reach of traditional perturbative techniques. Adding to this difficulty is the wide range of physical scales involved: the dynamics span energies from the sub-MeV domain relevant to binding energies, through the muon mass scale, and potentially beyond the electroweak scale in the presence of heavy new physics. Moreover, the emission of energetic radiation introduces large logarithms that must be resummed to achieve accurate predictions. A complete understanding of spectral reshaping QED corrections thus requires a framework that consistently handles both the bound-state dynamics and the associated multiscale structure.

The natural framework is that of an effective field theory (EFT), as it allows an organized separation of different scales. Above the nuclear scale, the challenges described earlier do not arise; the relevant physics is perturbative and can be handled using standard techniques. At the nuclear scale, the problem becomes intertwined with nonperturbative nuclear physics, which governs the interaction of the muon and electron with the finite-size nucleus. While these effects are highly relevant for precise predictions, they are conceptually distinct from the QED corrections and, in the first approximation, do not affect the shape of the spectra. An EFT treatment extending down to and including the nuclear scale has recently been discussed in refs.~\cite{Haxton:2022piv,Haxton:2024lyc}.

Below the nuclear scale, the complexities associated with QED corrections in bound states truly emerge. This regime still involves a rich hierarchy of scales, most notably the cutoff energy $\Delta E$ below which soft photons remain undetected. The hierarchy can be illustrated with muon conversion. Working in the rest frame of the incoming nucleus, it begins with the muon mass, $m_{\mu} \sim 100 \, \mathrm{MeV}$, of the same order of the outgoing electron energy, $E_e$. Below this lies the scale of the muon’s momentum, $|\vec{p}| \sim Z \alpha m_{\mu}$, around $10 \, \mathrm{MeV}$ in the case of aluminum ($Z=13$). Further down is the electron mass, $m_e \sim 0.5 \, \mathrm{MeV}$, which not only is of the order of the scale of the muon binding energy, $E_b \sim (Z \alpha)^2 m_{\mu}$, but also of the experimental resolution scale $\Delta E$. Finally, photon emissions from the outgoing electron induce an even lower scale, $m_e \Delta E / m_{\mu} \sim 1 \, \textrm{keV}$. In sum, 
\ali{
&m_\mu \sim E_e \gg Z \alpha m_{\mu} \gg (Z \alpha)^2 m_{\mu} \sim m_e \sim \Delta E  \gg m_e \frac{\Delta E}{m_{\mu}}. \nonumber 
}

In this Letter, we introduce an EFT framework that systematically includes QED corrections across this hierarchy of scales. Equally important, the framework is not restricted to muon conversion but applies also to DIO, thereby addressing the challenges posed by the two key processes relevant for upcoming muon conversion searches. Our EFT framework is in fact much more general, extending to a wide array of bound-muon decays, including muon-to-positron conversion \cite{Lee:2021hnx}, radiative muon decay \cite{Uesaka:2024tfn}, and $X$-emitting muon decay (with $X$ being a neutral boson) \cite{Uesaka:2020okd}. This generality makes the framework a powerful tool for new physics searches beyond muon conversion: if deviations are observed in any of these channels, precise theoretical predictions for the spectrum shape will be essential.

The generality is enabled by modern EFTs, whose universality permits a unified treatment of these distinct processes. Specifically, our framework combines in a non-trivial way several modern EFTs:
Heavy Quark Effective Theory (HQET) \cite{Isgur:1989vq, Isgur:1990yhj, Neubert:1993mb, Manohar:1997qy, Manohar:2000dt}, 
Non-Relativistic QED (NRQED) \cite{Caswell:1985ui, Kinoshita:1995mt, Paz:2015uga},
potential NRQED (pNRQED), \cite{Pineda:1997bj, Pineda:1997ie, Brambilla:1999xf, Beneke:1999qg, Beneke:1998jj},
Soft-Collinear Effective Theory (SCET) \cite{Bauer:2000ew, Bauer:2000yr, Bauer:2001ct, Bauer:2001yt, Beneke:2002ph, Beneke:2002ni}
and boosted HQET (bHQET) \cite{Fleming:2007qr, Fleming:2007xt}.
These techniques have been recently applied to low-energy processes involving widely separated scales \cite{Hill:2016gdf,Beneke:2017vpq,Beneke:2019slt, Beneke:2020vnb, Beneke:2021jhp, Beneke:2022msp,Tomalak:2022xup, Tomalak:2022kjd, Hill:2023bfh, Hill:2023acw, Tomalak:2024lme, Cirigliano:2024msg,Cirigliano:2024rfk}. They were originally developed in the context of QCD, where nonperturbative effects often obstruct complete analytic control, requiring input from experiment or lattice QCD. One novelty in our approach lies in the application of those techniques to QED corrections in bound muons, where the dynamics remain perturbative, enabling fully controlled, high-precision calculations.

This Letter offers a concise presentation of the framework, aimed at making its structure and implications transparent for a broad range of applications. We expect it to serve as a foundation for future precision studies and new physics searches.
We illustrate the framework's usefulness by computing the most precise shapes to date for the spectra of both muon conversion and DIO. In the case of conversion, the framework complements the aforementioned refs.~\cite{Haxton:2022piv,Haxton:2024lyc}, by providing an EFT treatment below the nuclear scale. This supplies a crucial ingredient for establishing a comprehensive connection between new physics models and experimental results.

\vspace{2mm}
\noindent
\textbf{Effective field theory.}
Bound-muon decays have been studied using one of two approaches: either numerical methods, which are exact in the muon velocity $v$ but restricted to leading order in $\alpha$ \cite{Uberall:1960zz,Haenggi:1974hp,Shanker:1981mi,Watanabe:1987su,Watanabe:1993emp,Shanker:1996rz,Czarnecki:2011mx,Heeck:2021adh,Kaygorodov:2025yag}, or analytical expansions in $v$, which yield closed-form expressions but rely on the non-relativistic limit $v \ll 1$ \cite{Szafron:2015kja}.

The EFT framework offers several decisive advantages. First, it is systematically improvable: both the operator basis and the Lagrangian admit a controlled expansion in small parameters, enabling homogeneous power counting and consistent inclusion of higher-order corrections. This provides a rigorous handle on theoretical uncertainties, absent in previous approaches.
Second, EFTs enforce a clean separation of physical effects at different energy scales, avoiding double counting and enabling a modular structure in computations. Third, integrating out non-dynamical modes resolves large logarithms that would otherwise spoil perturbative convergence. This naturally yields all-orders factorization theorems, organizing contributions of various modes through renormalization-group evolution.
The EFT framework is particularly well-suited to bound-state problems, where it provides a quantum field theory (QFT) underpinning for nonrelativistic constructs such as the Schr\"odinger wavefunction. In contrast, the approaches mentioned above lack a systematic justification for borrowing ingredients from perturbation theory developed for scattering states in order to describe real emissions in bound systems, and they offer no clear prescription for separating overlapping regimes or resumming large logarithms.

As already suggested, the universality of the EFT framework is equally powerful. Provided the outgoing charged lepton carries energy close to the muon mass, bound-muon decays as different as muon DIO and muon-to-positron conversion are governed by the same momentum regions \cite{Beneke:1997zp,Smirnov:1990rz,Ma:2023hrt}.  While each channel features process-specific functions, the majority of ingredients is shared. This makes the EFT framework remarkably versatile: once the universal building blocks are established, QED effects can be systematically and efficiently computed across a whole class of processes. 

In what follows, we outline the essential features of the framework. It comprises a sequence of EFTs, one for each relevant scale; details can be found in refs. \cite{Fontes:2024yvw,Fontes:2025mps}.
To illustrate phenomenological applications, we consider muonic aluminum, which has been chosen for upcoming experiments \cite{Mu2e:2014fns,COMET:2018auw}. 
Other elements are of interest as well, and our framework applies directly to nuclei with atomic numbers around 10. For significantly heavier or lighter nuclei, the framework can be systematically extended, by reshuffling contributions to account for the altered scale hierarchies.

We define the power-counting expansion parameter
\ali{
\label{eq:lambda}
\lambda \sim Z \alpha \sim \sqrt{\dfrac{m_e}{m_{\mu}}}.
}
Given the complexity of the framework, we restrict our analysis to the leading power (LP). However, the formalism we develop is systematically improvable and allows for the inclusion of subleading power corrections, which are expected to prove phenomenologically relevant in planned high-precision future experiments \cite{Kuno:2005mm,Mu2e-II:2022blh,CGroup:2022tli}.

\textit{Hard scale, $\mu_{\hvar} \! \sim m_{\mu}$}. This marks the starting point of our EFT analysis. Since it lies far below the nuclear mass $M_N$ ($\sim 25\,\text{GeV}$), the nucleus as a dynamical particle is integrated out. Nuclear recoil can be systematically organized in terms of a parameter $\lambda_R \sim m_{\mu}/M_N$. On the other hand, the nuclear radius $r_n$ satisfies $r_n \sim 1/\mu_{\hvar}$. Finite-nucleus-size (FNS) effects therefore induce a non-trivial structure in the amplitude at the hard scale, encoding nonperturbative nuclear physics. These effects are discussed in the Appendix. For the leptons, the theory at the hard scale is that of weak interactions below the electroweak scale. It follows that the muon and the electron are described by the QED Lagrangian, with the electron mass contributing only at subleading power.

\textit{Semi-hard scale, $\mu_{\shvar} \sim \lambda m_{\mu}$}. Fluctuations of $\mathcal{O}(m_{\mu})$ are now too rapid and are integrated out.
The nucleus is described by local fields within HQET, such that FNS effects are codified in the EFT coefficients.
For the muon, a non-relativistic treatment is required, more specifically within NRQED. The photon-mediated interactions with the nucleus remain perturbative here, since the muon has not yet formed a bound state.
For the outgoing electron, while at the hard scale all interactions were of the same order and could be described within local QED, at the semi-hard scale the electron exhibits a hierarchy of modes: it retains an energetic component along its direction of motion while simultaneously interacting with softer radiation. This separation of scales would render a QED approach inconsistent, and requires the more refined, non-local description provided by $\rm SCET_{I}$.

\textit{Soft scale, $\mu_{\svar} \! \sim \! m_e$}. The photon-mediated interactions between the muon and the nucleus can no longer be treated perturbatively, and the repeated exchange of potential photons leads to the formation of a bound state. The appropriate EFT is pNRQED, where these nonperturbative effects are encoded in instantaneous potentials $V$ between the potential muon field $\Psi^{\p}$ and the static nucleus field $h_N^{\s}$. While at LO the potentials reduce to the Coulomb interaction, the exchange of genuinely radiative photons remains perturbative and is included as higher-order corrections to the bound-state dynamics. The inclusion of $V$ in the Green's function leads to the appearance of the bound-state wavefunction; more specifically, since bound-muon decays probe short-distance physics relative to the atomic scale, the leading approximation depends only on the wavefunction at the origin.
The latter is modified by FNS effects; in the EFT framework, these corrections are divergent and require a careful treatment, detailed in the Appendix. There, we also show that they induce moderate numerical corrections ($\sim \! 5\%$), which is consistent with the fact that power counting qualifies such FNS effects as next-to-next-to-leading power (NNLP) in the $\lambda$ expansion.
Finally, the electron mass at the soft scale is no longer suppressed; this gives rise to a more involved soft-collinear description for the outgoing electron, namely that of $\rm SCET_{II}$.

\textit{Soft-collinear scale, $\mu_{\scvar} \! \sim \! m_e \Delta E/m_{\mu}$}. Whereas in the simple bound-muon problem the analysis terminates at the soft scale, the presence of the energetic outgoing electron forces us to descend further to $\mu_{\scvar}$. Since $m_e \gg \mu_{\scvar}$, it is no longer possible to virtually generate soft electrons. The collinear fluctuations of the outgoing electron, still carrying large energy, are now too rapid and must be integrated out. Its energy can then fluctuate only within a small range of order $\Delta E$, and the appropriate description is in terms of a boosted HQET (bHQET) field $h_{e}^{\sc}$. After this, the LP Lagrangian for muon conversion reads
\begin{widetext}
\vspace{-5mm}
\ali{
\label{eq:Lag}
\mathcal{L} =
\bar{h}_{N}^{\s} i \partial_0 h_{N}^{\s}
+ 
\bar{\Psi}^{\p} \left( i \partial_0 + \dfrac{\vec{\nabla}^2}{2 m_{\mu}} \right) \Psi^{\p} + \int d^3 r \, \bar{h}_{N}^{\s}\ar{x} h_{N}^{\s}\ar{x} V\ar{\vec{r}} \bar{\Psi}^{\p}\ar{x+\vec{r}} \Psi^{\p}\ar{x+\vec{r}}
+
\bar{h}_{e}^{\sc} i v_e \cdot \partial h_{e}^{\sc},
}
\end{widetext}
with $v_e=(m_e/(4 m_{\mu}) + m_{\mu}/m_e, 0, 0, m_e/(4 m_{\mu}) - m_{\mu}/m_e)$ and where all the fields have been decoupled from interactions with both soft and soft-collinear photons. Factorization is thus achieved for bare LP quantities: all the relevant modes for muon conversion and DIO are completely separated at the level of the LP Lagrangian. It is worth stressing that a possible infinite cascade of modes due to the massive outgoing electron is tamed by the measurement function.

\vspace{2mm}
\noindent \textbf{Factorization.}
Having derived the Lagrangian of the final EFT, we may formulate a factorization theorem, which is valid to all orders in $\alpha$ at LP in $\lambda$, and which organizes the contributions from the relevant modes. The resulting factorization formul\ae \, for the multiple bound-muon decays discussed before are structurally similar, highlighting the universality of the EFT framework. In fact, the normalized cumulant distribution can be written in a generic way as
\begin{widetext}
\vspace{-5mm}
\ali{
\frac{\Gamma_{\rm{cum}}(\Delta E)}{\Gamma^{\rm{LO}}} \equiv \hspace{-5mm}
\int\limits_{E_e^{\rm max} - \Delta E}^{E_e^{\rm max}}
\hspace{-4mm}
\frac{dE'_e}{\Gamma^{\rm{LO}}}
\frac{d \Gamma}{dE'_e}
\!=\!
|\psi_{\rm corr}|^2  |\mathcal{H}|^2 |C_m|^2 |U_{\hvar}|^2 \! \int \! dE_{\scvar} \, dE_{\svar} \,
\mathcal{S}(E_{\svar}) \mathcal{SC}(E_{\scvar}) (\Delta E \!-\! E_{\svar} \!-\! E_{\scvar})^N \theta(\Delta E \!-\! E_{\scvar} \!-\! E_{\svar}),
}
\end{widetext}
with $N$ being a natural number that depends on the process and $E_e^{\rm max}$ the endpoint energy of the outgoing electron.
The cumulant is introduced to provide a generic formula that applies to a broad class of bound-muon decays. 
For presentation, we normalize to $\Gamma^{\rm LO}$; although other normalizations could be adopted, this choice is of secondary importance since our focus lies on the shape.
Each function in the factorization formula is associated with a single scale: the hard function $\mathcal{H}$ is evaluated at the hard scale, the wavefunction correction $\psi_{\rm corr}$ at the semi-hard scale, the radiative jet function $C_m$ at the soft scale, and the soft and soft-collinear functions, $\mathcal{S}$ and $\mathcal{SC}$, at the soft and soft-collinear scales, respectively. The formula also includes the evolution factor $U_{\hvar}$, which accounts for the running between the hard and soft scales.

The universality of the EFT approach is now evident: across the wide range of bound-muon decay processes considered, the only process-dependent ingredients are the exponent $N$ (e.g. $N=6$ for DIO and $N=0$ for conversion) and the hard function along with its renormalization-group evolution. This structural simplicity makes the calculation of QED corrections in bound-muon decays remarkably efficient.

We choose the soft scale as the common scale to which all matching coefficients are evolved. This choice significantly simplifies the factorization structure for three reasons. First, two of the functions, $C_m$ and $\mathcal{S}$, are naturally evaluated at the soft scale; evolving all coefficients to this scale eliminates the need for their corresponding evolution factors. Second, a consistent treatment of the evolution of $C_m$ and  $\mathcal{S}$ requires the use of the rapidity renormalization group. Third, the running of the soft-collinear function $\mathcal{SC}$ from the soft-collinear scale up to the soft scale is fully controlled by abelian exponentiation; below the soft scale, the theory contains only photons, and electron loops are absent.

The functions $\mathcal{H}$, $\psi_{\rm corr}$ and $C_m$ arise from the sequence of EFTs outlined above. More precisely, they are determined through matching at successive scales: between the hard and semi-hard, the semi-hard and soft, and the soft and soft-collinear scales, respectively. 
The soft and soft-collinear functions enter the factorization formula as a convolution, reflecting the interplay between emissions at different momentum regions. These functions originate from the separation of real radiation into two distinct modes: wide-angle soft radiation and collinear radiation with soft energy. This bipartition is essential for capturing the correct infrared (IR) structure of the process. Importantly, $\mathcal{S}$ and $\mathcal{SC}$ are the functions responsible for distorting the shape of the conversion spectrum: in their absence, the LO prediction for the cumulant distribution would be flat, and no structure would emerge near the endpoint. The derivation of all these functions, along with the relevant corresponding evolution factors, is presented in detail elsewhere \cite{Fontes:2024yvw,Fontes:2025mps}. 

Generic real collinear radiation is vetoed by the requirement that the electron energy in near the endpoint, so the collinear sector contributes only virtually. Accordingly, the “jet function” is an exclusive object that coincides with the matching coefficient $C_m$ when integrating out collinear modes at the soft/soft-collinear threshold, i.e. when matching the SCET collinear electron onto a bHQET electron field. The coefficient $C_m$ resums the associated jet logarithms, while real emissions are encoded in the soft and soft-collinear functions. This structure mirrors known exclusive SCET factorizations, but here it is implemented for bound-muon conversion with a massive electron, which to our knowledge has not been formulated before in this context.

\begin{figure}[b!]
\centering
\includegraphics[width=0.49\textwidth]{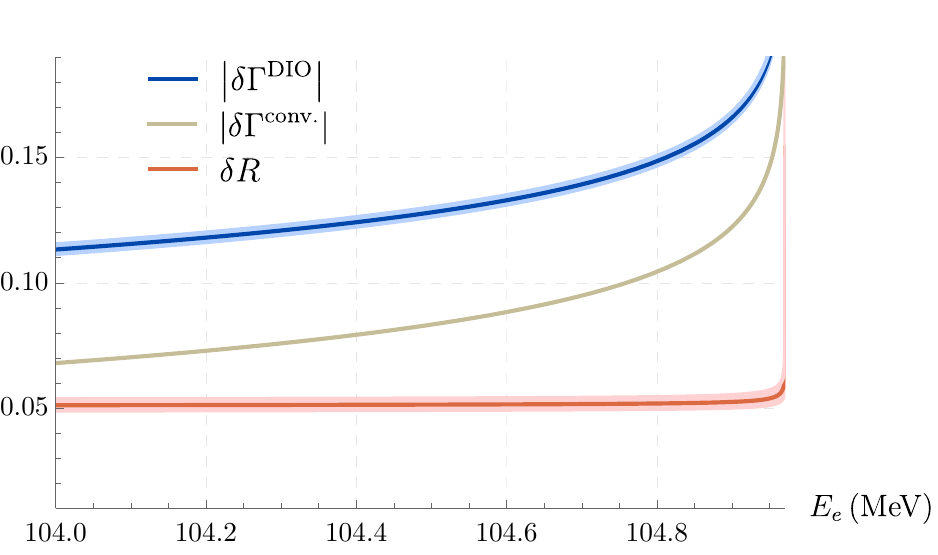}
\caption{Normalized cumulant distributions for muon DIO, conversion and their ratio in the endpoint region of the electron spectrum. While the DIO and conversion curves show sizable, energy-dependent corrections exceeding $10\%$, their ratio remains $5\%$ and exhibits only a very weak energy dependence. See text for details.}
\label{fig:zoom}
\end{figure}

\vspace{2mm}
\noindent
\textbf{Impact on the shape of the rates.}
As an application of our framework, we compute the shapes of both muon conversion and DIO. Our analysis follows the standard logarithmic hierarchy of QCD, with our most precise predictions being obtained with next-to-leading-logarithmic-prime resummation (NLL'). We define
\ali{
\delta \Gamma^{\rm{P}} \equiv \frac{\Gamma_{\rm{cum,NLL'}}^{\rm{P}} - \Gamma_{\rm{cum,LO}}^{\rm{P}}}{\Gamma_{\rm{cum,LO}}^{\rm{P}}}
}
for a certain process P and
\ali{
\delta R \equiv \frac{R_{\rm{NLL'}}-R_{\rm{LO}}}{R_{\rm{LO}}},
\hspace{3mm}
\textrm{with}
\hspace{3mm}
R_{\textrm{X}} \equiv \frac{\Gamma_{\rm{cum,X}}^{\rm{conv.}}}{\Gamma_{\rm{cum,X}}^{\rm{DIO}}},
}
where we omitted the argument $\Delta E$ for compactness.
Figure~\ref{fig:zoom} shows $\delta \Gamma^{\rm P}$ for muon DIO (blue) and conversion (beige), as well as $\delta R$ (red), against $E_e \equiv E_e^{\rm max} - \Delta E$. All curves include a standard 7-point scale variation, obtained by varying the scales $\mu_{\hvar}$ and $\mu_{\svar}$ around the respective central values $2 m_{\mu}$ and $m_e$ by factors of $1/2$ and $2$, with the two extreme combinations omitted (the envelope in $\delta \Gamma^{\rm{conv.}}$ is negligible).  
The plot shows that QED corrections to muon DIO and conversion vary in a non-trivial way with the distance to the endpoint and can clearly exceed $10\%$. In contrast, when conversion is normalized to DIO as in $\delta R$, the corrections are essentially $5\%$ and constant throughout the endpoint region. Moreover, the FNS dependence of the wavefunction at the origin cancels. 

Overall, the figure shows that QED corrections to the shape are sizable and must be brought under control in order to enable a reliable interpretation of upcoming experimental results. The framework developed here provides this control and therefore represents a key element for precision studies of bound-muon decays.

\vspace{2mm}
\noindent
\textbf{Discussion.}
In this Letter, we have presented a consistent and systematic framework for computing higher-order corrections to bound-muon decays below the nuclear scale. A defining feature of our approach is its universality: by leveraging the shared dynamical structure of processes with energetic final-state leptons, our framework applies to a wide class of processes, not just a single channel. It provides proper QFT definitions, a homogeneous power counting, Feynman rules suited for multi-scale dynamics, and the resummation of large logarithms, integrating tools from HQET, NRQED, SCET and related EFTs. As an illustration, we computed at LP and at NLL' in QED the spectral shapes of both muon conversion and decay-in-orbit. These processes are central to upcoming searches, and we find that QED corrections can modify the signal shape by more than $10\%$, a substantial effect with direct impact on discovery potential. With future experiments aiming for unprecedented sensitivities, such theoretical precision may prove decisive in enabling or excluding the discovery of new physics across multiple channels. Our framework also represents a crucial step toward connecting high-scale physics to low-energy observables with theoretical control, complementing recent progress above the nuclear scale and completing the EFT bridge down to experimental energies.

\vspace{2mm}
\noindent
\textbf{Acknowledgments}.
All Feynman diagrams were drawn with Feyngame \cite{Harlander:2020cyh, Harlander:2024qbn,Bundgen:2025utt}, and FeynCalc \cite{Mertig:1990an,Shtabovenko:2016sxi,Shtabovenko:2020gxv} was very useful.
D.F. thanks Mainz Institute for Theoretical Physics (MITP) of the Cluster of Excellence PRISMA+ (Project ID 390831469), as well as the Laboratoire de Physique Théorique et Hautes Energies, for their hospitality and support.
The research of D.F. was supported by the Deutsche Forschungsgemeinschaft (DFG, German Research Foundation) under grant 396021762 - TRR 257, and that of R.S. by the U.S. Department of Energy under Grant Contract No. DE-SC0012704.

\appendix

\section{Finite nucleus size}

\noindent The aluminum nucleus root-mean-square radius  is \cite{DeVries:1987atn,Fricke:1992zza}
\ali{
\label{eq:radius-size}
	r_n \simeq 0.0155(2)\, \textrm{MeV}^{-1} \sim 1/\mu_{\hvar},
}
of the order of the typical inverse hard scale. 
This results in significant FNS effects both at and below the hard scale. Since the nuclear radius serves as a physical cutoff for certain IR divergences, the EFT expansion naturally involves additional signatures associated with the perturbative treatment of the FNS. Here, we start by discussing the cancellation of those singularities within the EFT framework, and we conclude with estimates for the various contributions.

\vspace{2mm}
\noindent
\textbf{Renormalization.}
We first demonstrate how the effective pNRQED operators induced by the finite-range potential  lead to singularities in the bound-state wavefunction. Subsequently, we verify their cancellation against corresponding singularities present in the hard function.

\textit{Wavefunction.}
At the semi-hard scale, $\mu_{\shvar} \ll \mu_{\hvar}$, the fluctuations of typical size corresponding to the FNS are integrated out. The static nucleus is described by a static HQET field, $h_N^{\sh}$, and the FNS effects are codified in Wilson coefficients. More specifically, the relevant terms in the nucleus Lagrangian are (cf.~ref.~\cite{Hill:2012rh}) 
\ali{
\label{eq:Lag}
\mathcal{L}_{h_N}^{\sh} \ni \bar{h}_N^{\sh} \! \left( i D_0^{\sh} \!-\! \frac{e \, c_D}{8 M_N^2} \vec{\nabla} \cdot \vec{E}^{\sh} \right) h_N^{\sh},
}
where $e$ is the electric charge, $\vec{E}$ the electric field, and $c_D$ the first relevant coefficient for FNS effects, given by
\ali{
c_D = -Z \left(1 + \frac{4}{3} M_N^2 r_n^2 \right) + \mathcal{O}(\alpha).
}
We note that the coefficient $c_D$ is conventionally normalized by the nucleus mass scale squared, $M_N^2$. This choice is not motivated by our power-counting; in fact, the FNS contribution to $c_D$ originates after integrating out the hard modes. That contribution is actually suppressed by the muon (rather than the nucleus) mass, since the aluminum nucleus radius turns out to be of the order of inverse muon mass.
We also note that $r_n^2 \vec{\nabla} \cdot \vec{E}^{\sh} \sim \lambda^2$, which follows from eq.~(\ref{eq:radius-size}) and $\vec{\nabla} \cdot \vec{E}^{\sh} \sim \mu_{\shvar}^2$. That is, the first FNS effects that contribute to the wavefunction at the origin are an NNLP correction. 
\begin{figure*}[!t]  
\centering
\includegraphics[width=0.135\textwidth]{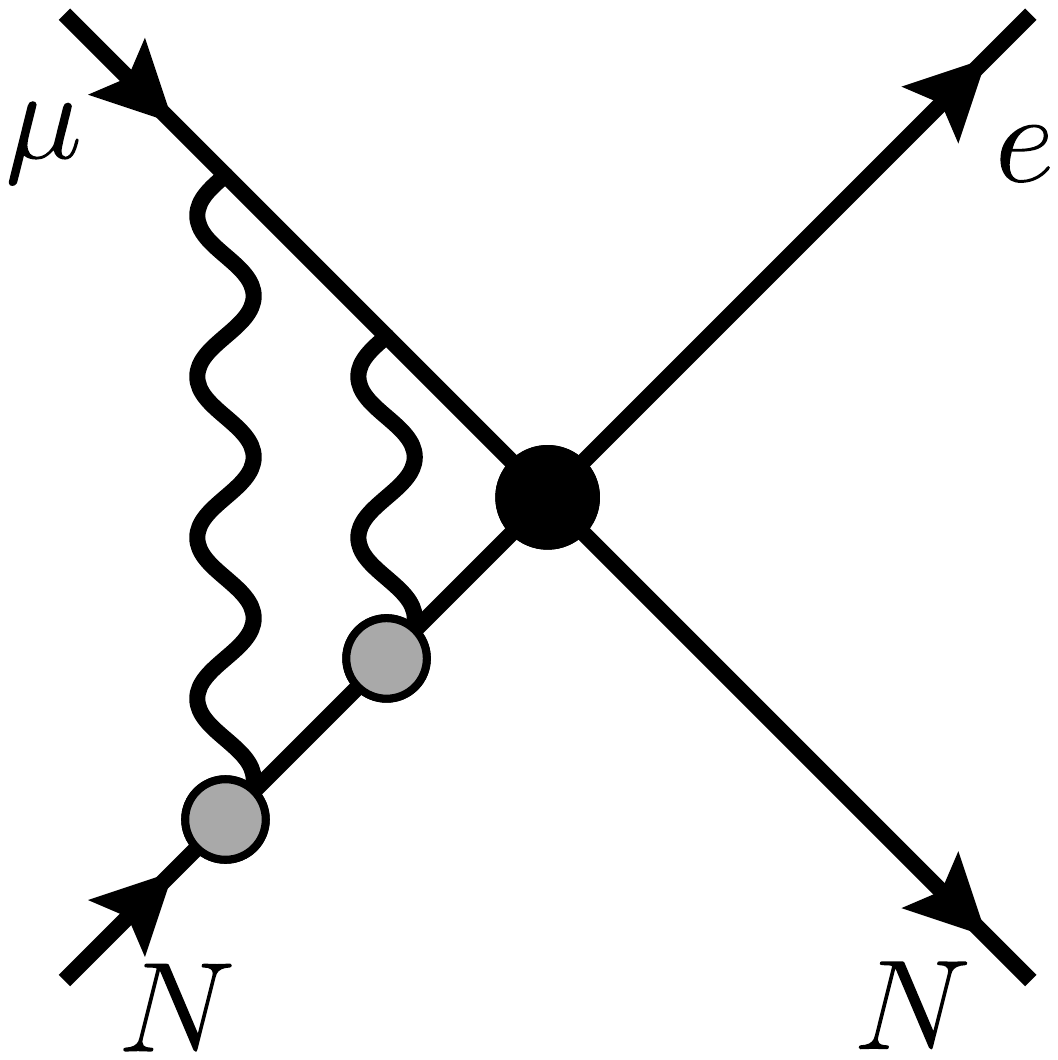}
\hspace{2.4mm}
\includegraphics[width=0.135\textwidth]{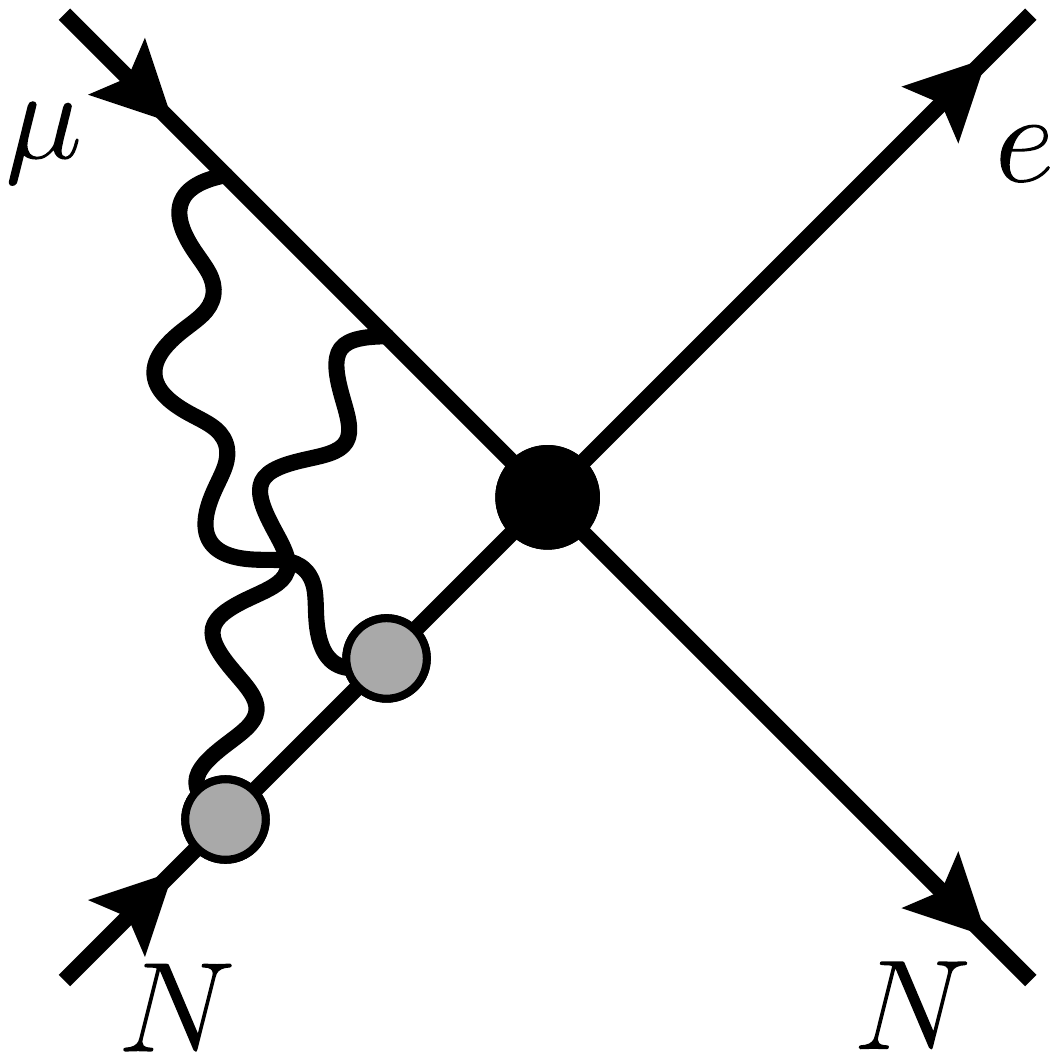}
\hspace{2.4mm}
\includegraphics[width=0.135\textwidth]{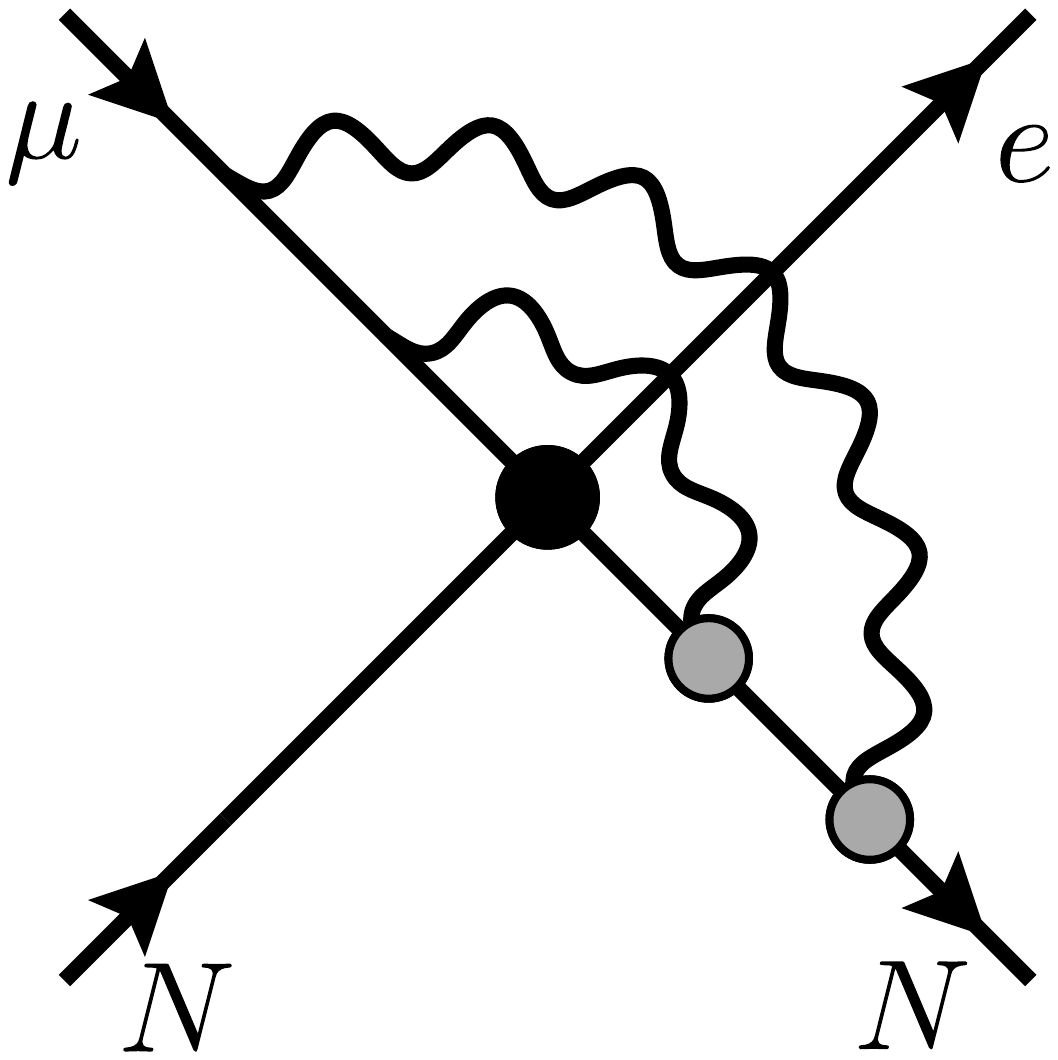}
\hspace{2.4mm}
\includegraphics[width=0.135\textwidth]{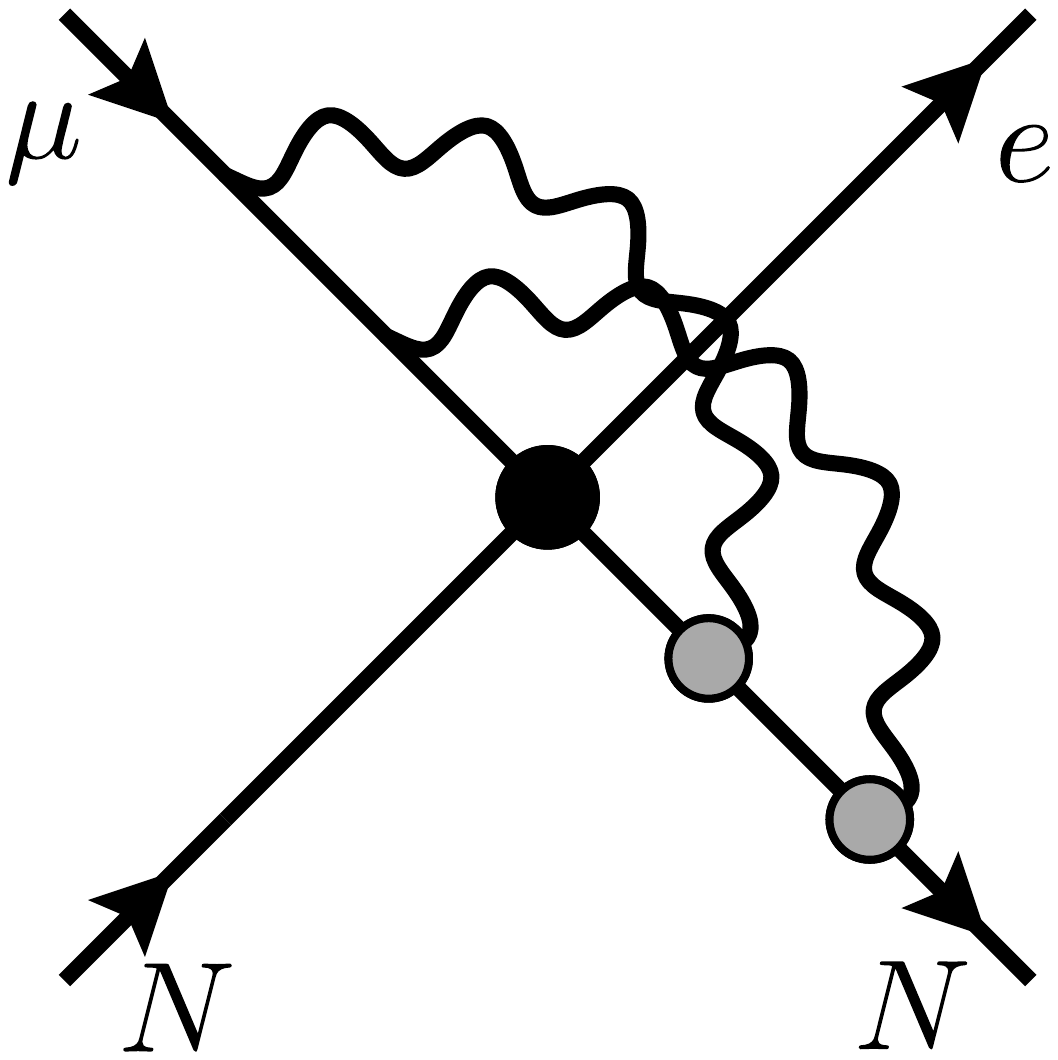}
\hspace{2.4mm}
\includegraphics[width=0.135\textwidth]{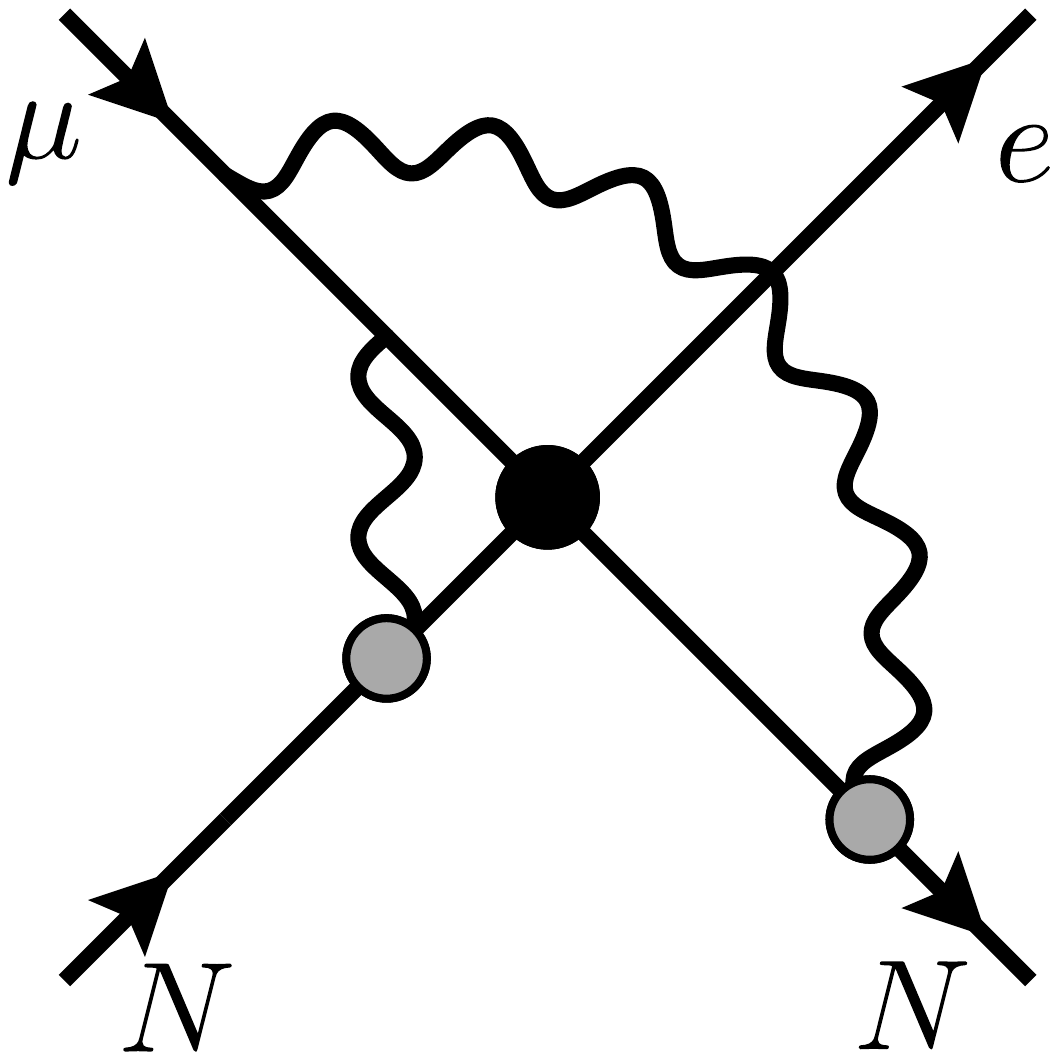}
\hspace{2.4mm}
\includegraphics[width=0.135\textwidth]{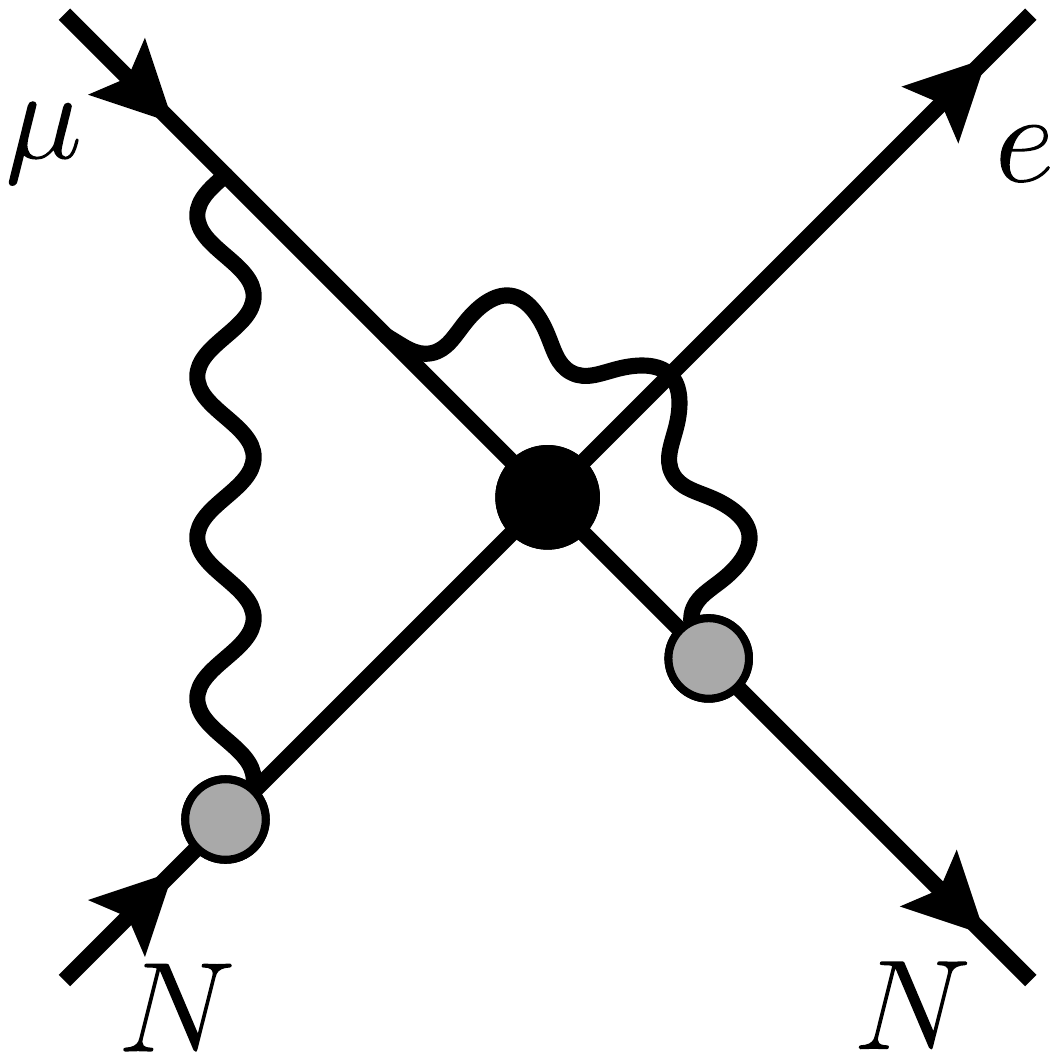} \\[-8mm]
\includegraphics[width=0.15\textwidth]{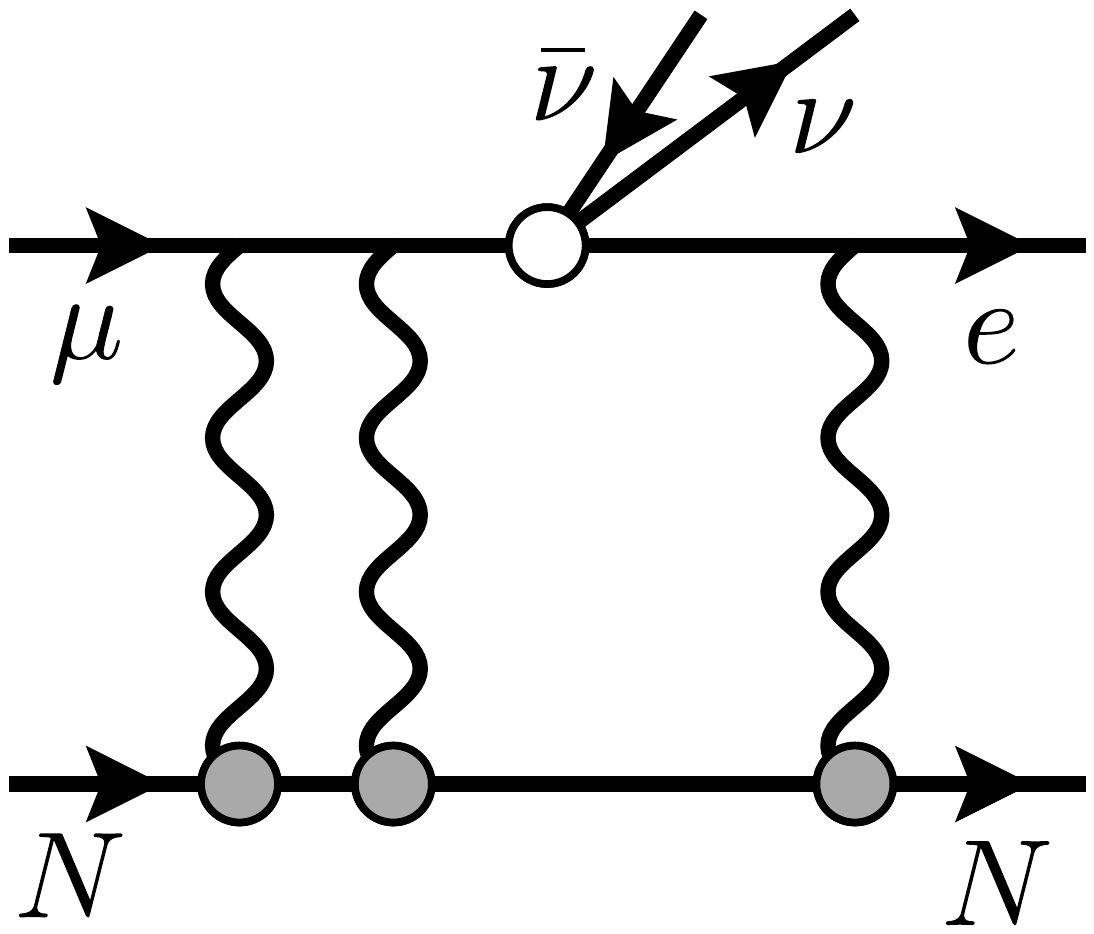}
\includegraphics[width=0.15\textwidth]{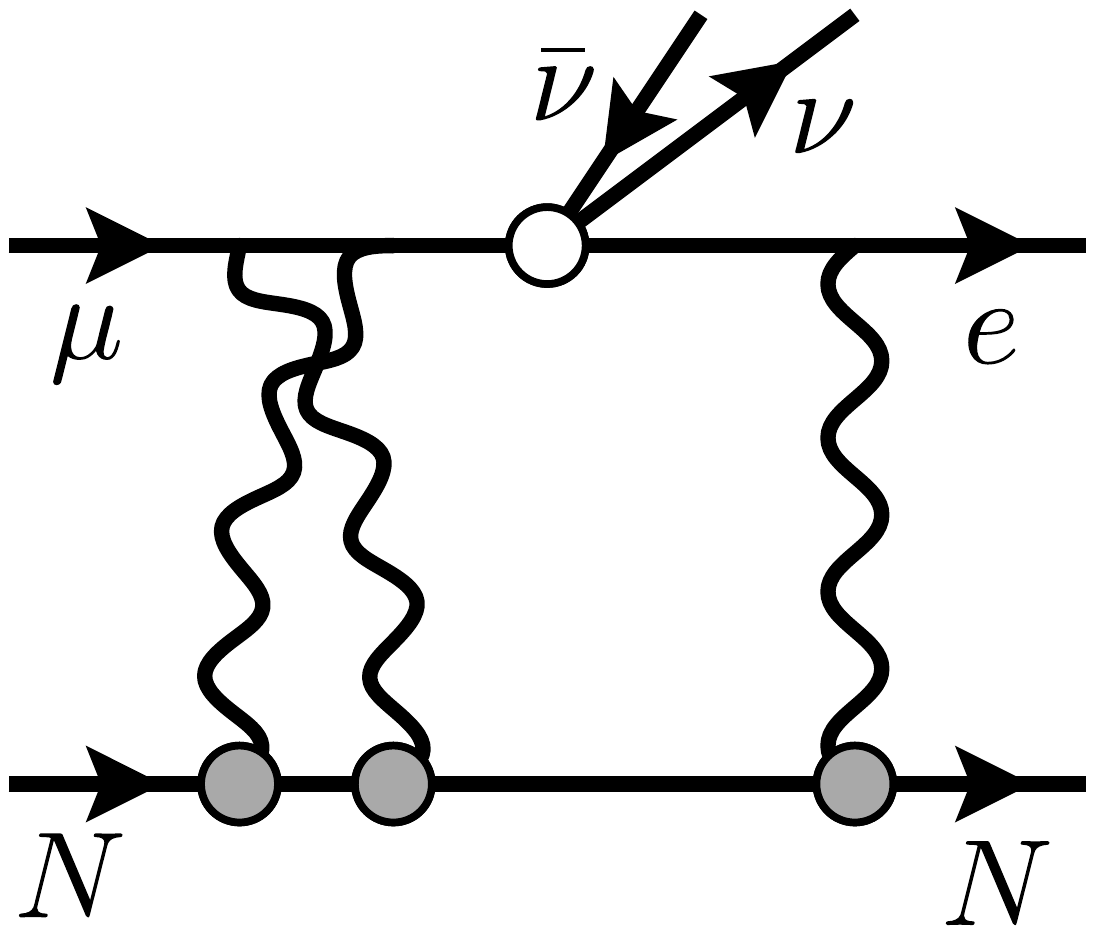}
\includegraphics[width=0.15\textwidth]{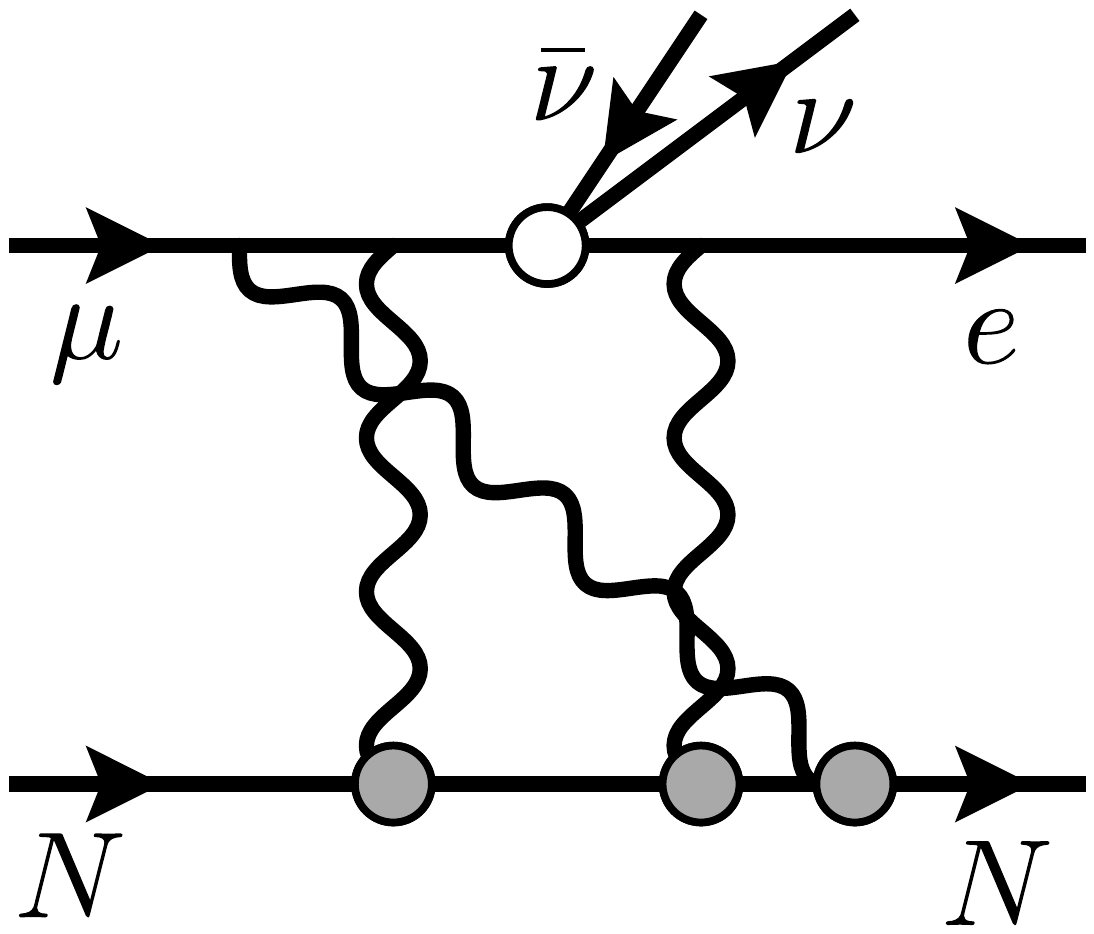}
\includegraphics[width=0.15\textwidth]{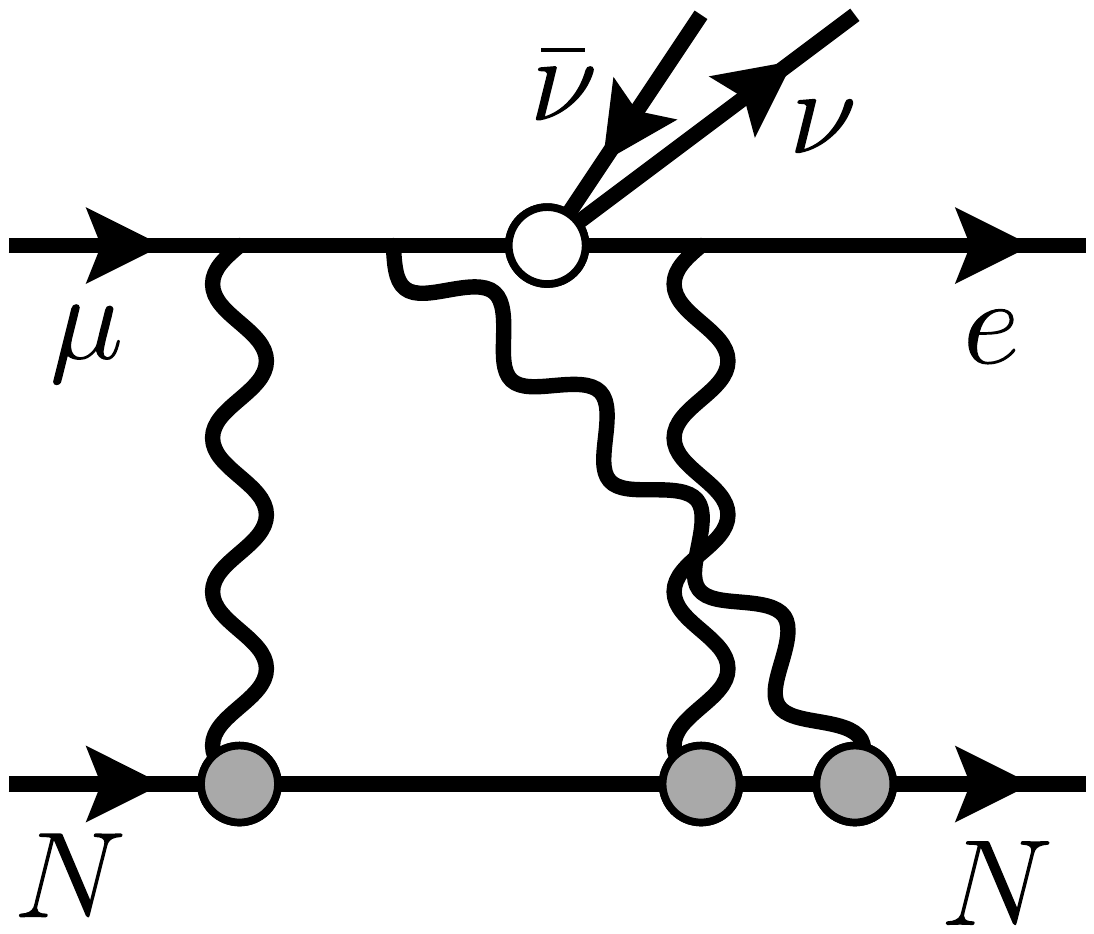}
\includegraphics[width=0.15\textwidth]{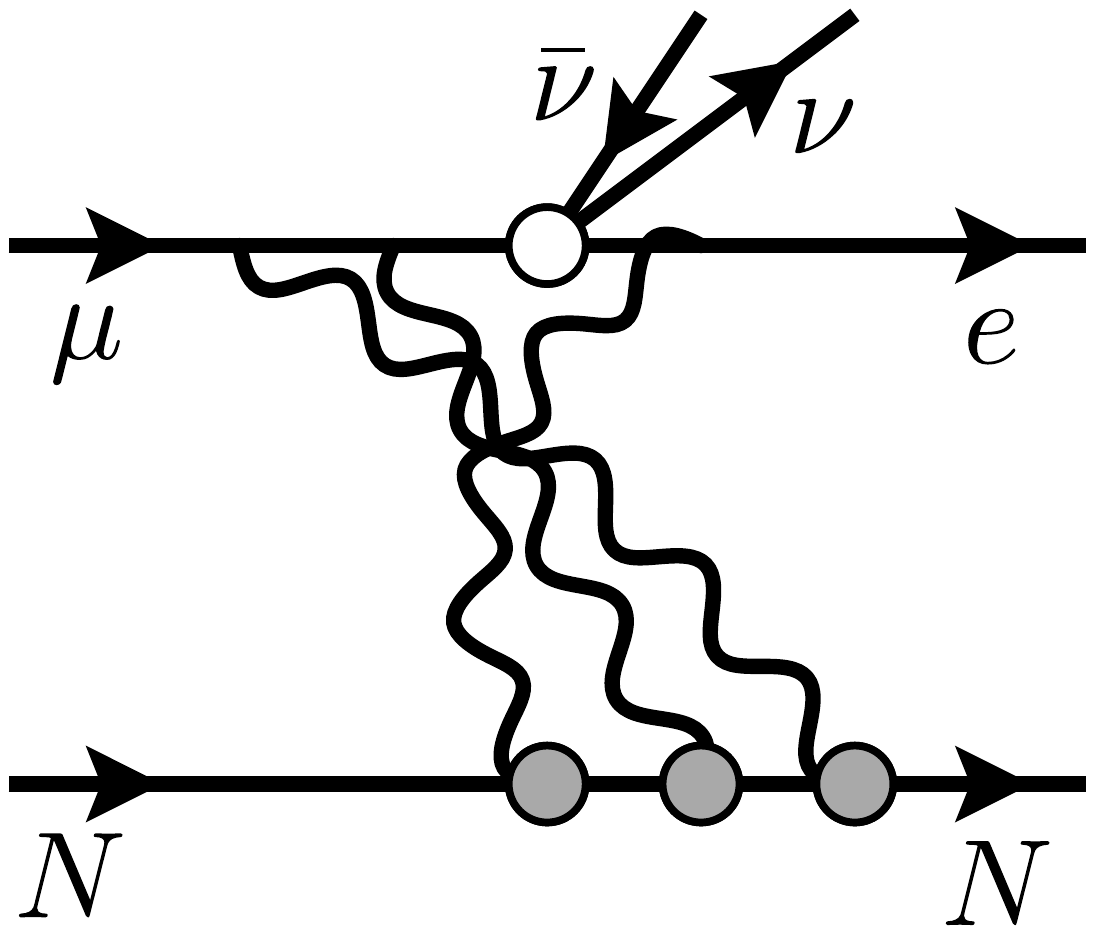}
\includegraphics[width=0.15\textwidth]{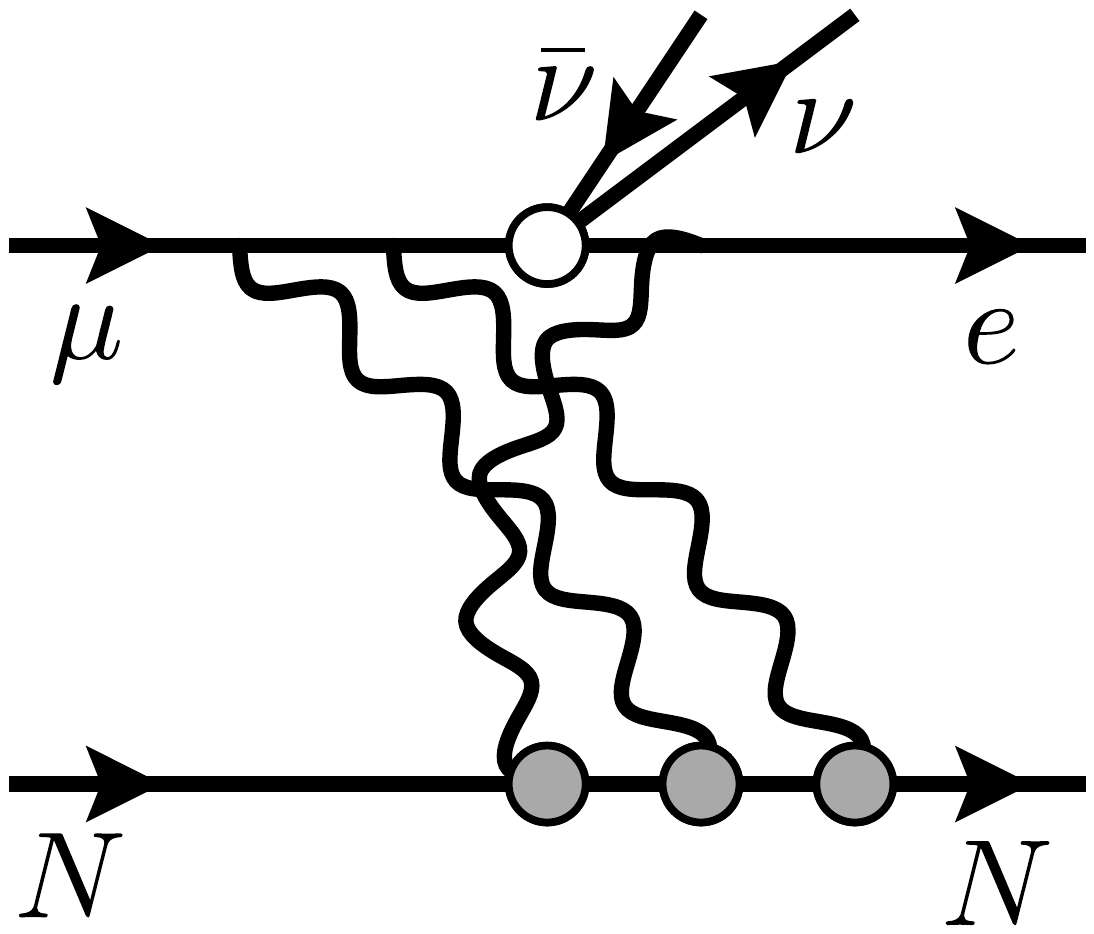} \\[-14mm]
\includegraphics[width=0.15\textwidth]{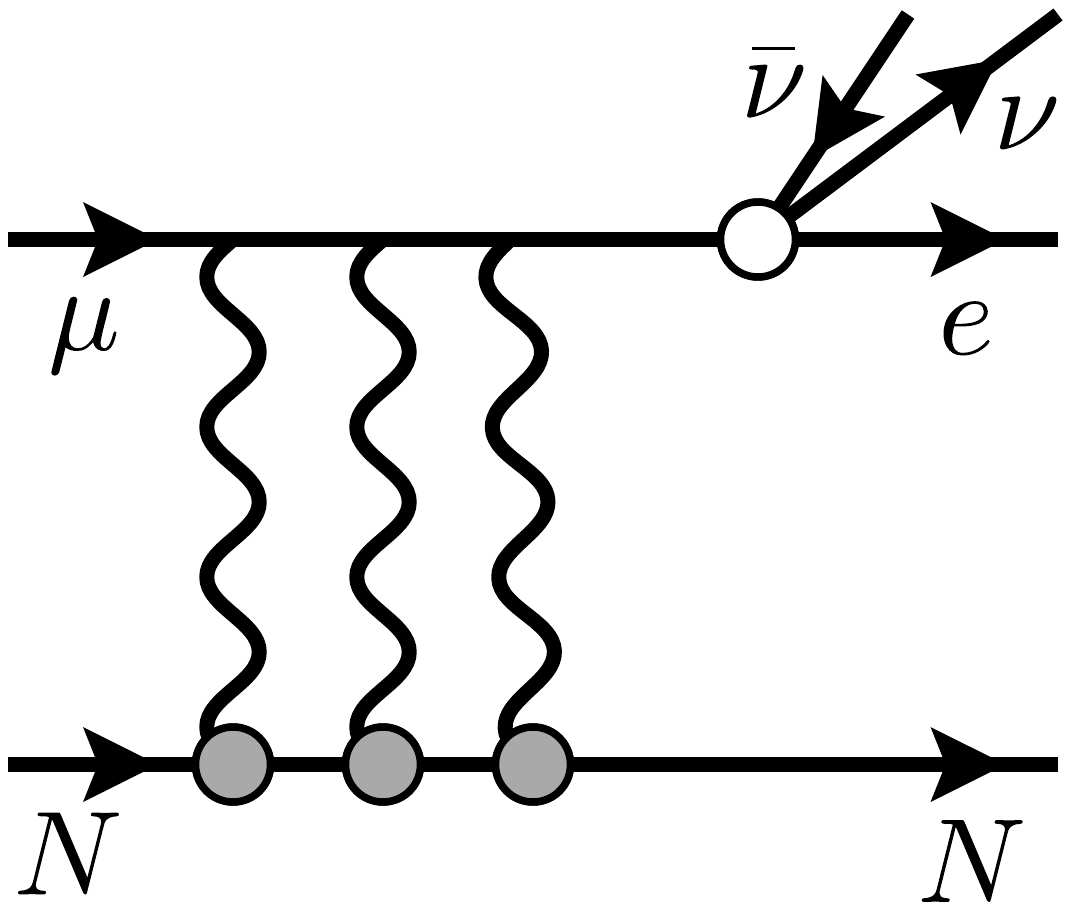}
\includegraphics[width=0.15\textwidth]{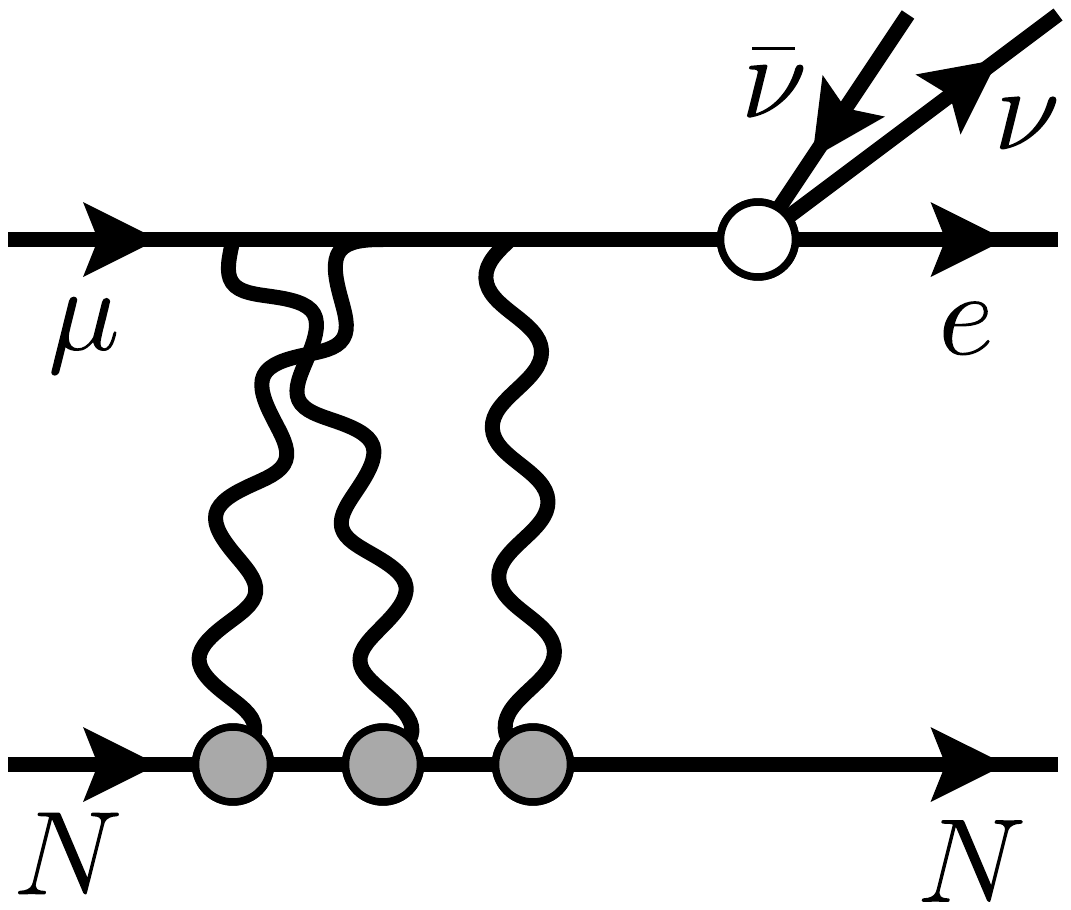}
\includegraphics[width=0.15\textwidth]{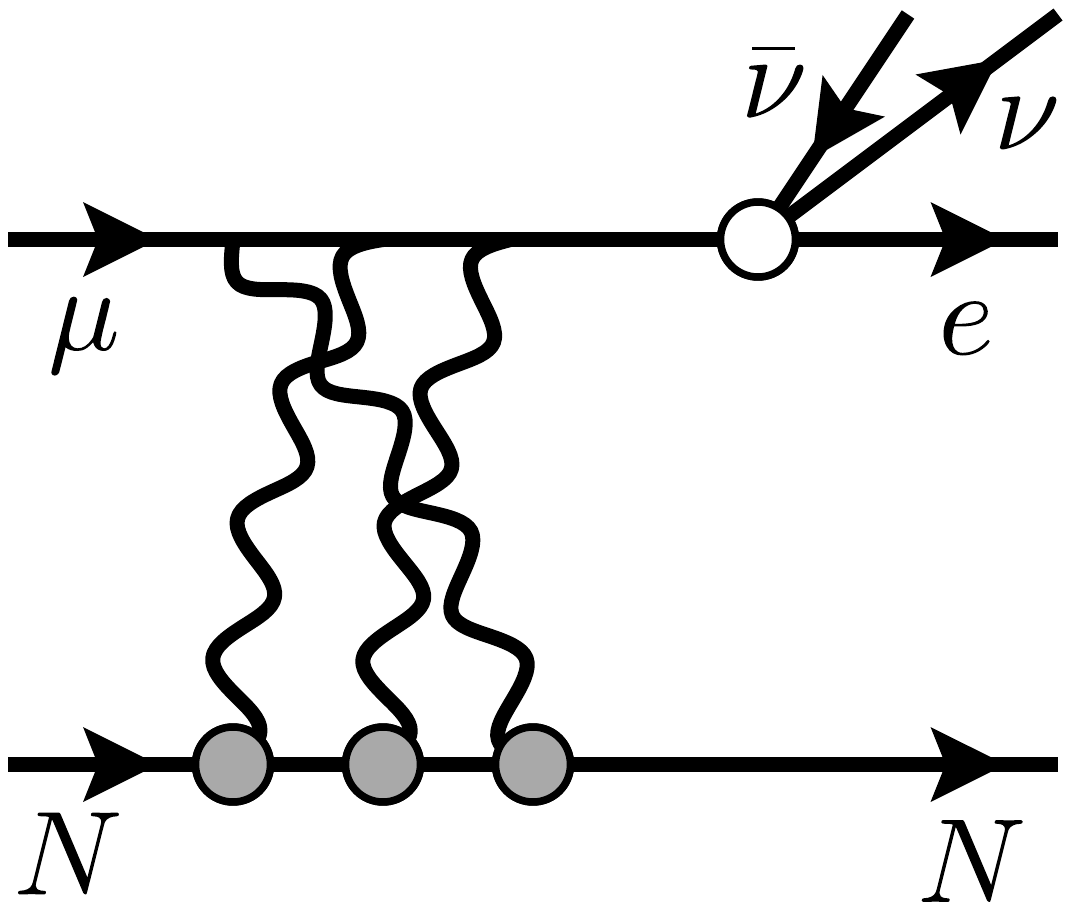}
\includegraphics[width=0.15\textwidth]{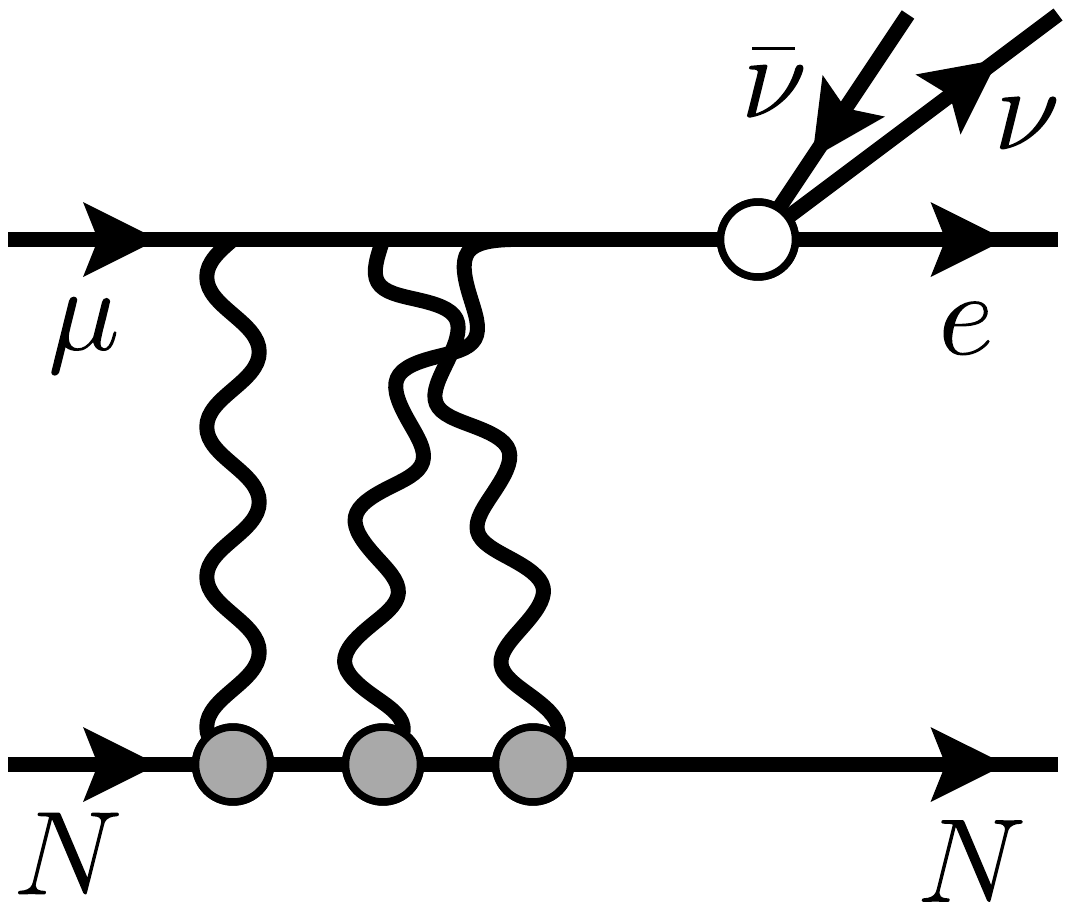}
\includegraphics[width=0.15\textwidth]{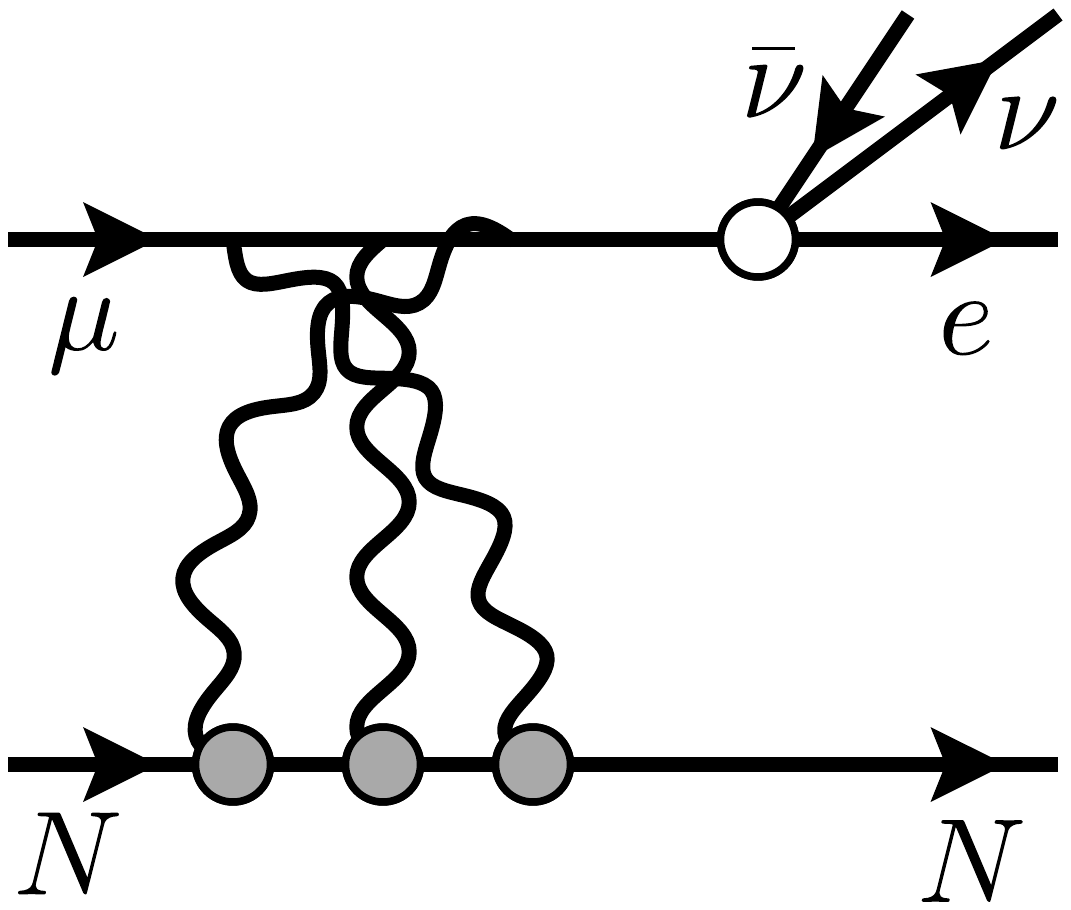}
\includegraphics[width=0.15\textwidth]{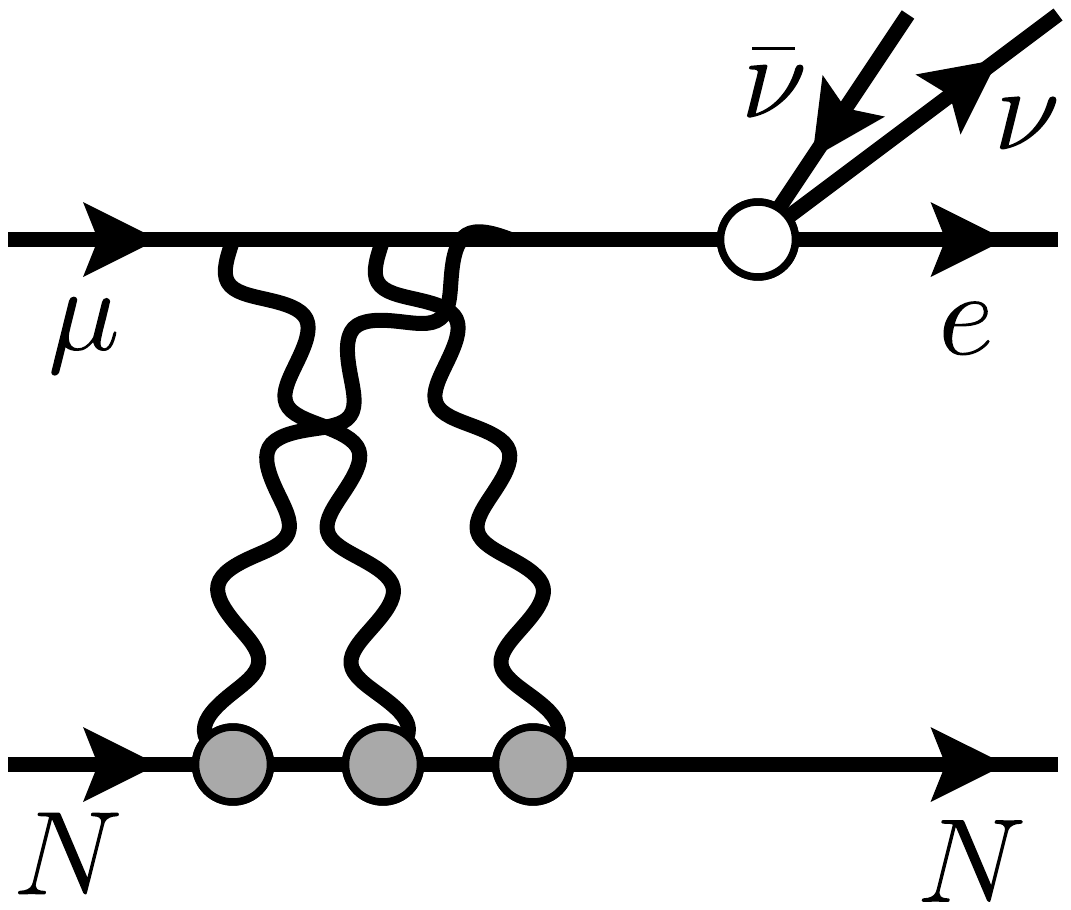}
\vspace{-15mm}
\caption{Feynman diagrams for the 2-loop photon-mediated interactions between muon and nucleus in muon conversion (first row) and DIO (second and third rows). The black disks represent the high-energy CLFV effective interaction, the white ones the Fermi interaction, 
and the gray ones the interaction between a photon and a non-point-like nucleus.}
\label{fig:diagrams}
\end{figure*}

Such correction comes about not at the semi-hard scale, but at the soft one. In the matching between the two scales, and denoting by $\vec{r}$ the relative coordinate between the muon and the nucleus, the LO potential is the Coulomb one, $V_{\rm LO}(\vec{r}) = -Z \alpha/|\vec{r}|$. Since $|\vec{r}| \sim 1/\mu_{\shvar} \sim 1/(Z \alpha m_{\mu})$, $V_{\rm LO}$ is not suppressed and must be considered to all orders as a nonperturbative effect. By contrast, the $r_n^2$ contribution to eq.~(\ref{eq:Lag}) is a perturbative correction to the potential,
$\delta V_{r_n^2} = Z e^2 r_n^2 \delta^{(3)}(\vec{r})/6 \sim (Z \alpha)^4 m_{\mu}$.

It is this correction that affects the wavefunction at the origin. From $H \psi = E \psi$ and by separating each quantity into a leading order (index 0) and a correction (index 1), we find $(H_0 - E_0) \psi_1 = (E_1 - H_1) \psi_0$. This implies $\psi_1 = (H_0 - E_0)^{-1} (E_1 - H_1) \psi_0$, with $(H_0 - E_0)^{-1}$ being
the Coulomb Green's function \cite{Schwinger:1964zzb,Beneke:2013jia}. That is,
\ali{
\psi_1(\vec{x}) = \int d^3x' G_0(\vec{x}, \vec{x}';E_0) (E_1 - H_1) \psi_0(\vec{x}') \, .
}
From $\delta V_{r_n^2}$ follows 
$H_1 = -\frac{e^2 c_D}{8 M_{N}^2} \delta^{(3)}(\vec{x}')$ and $E_1 =  -\frac{e^2 c_D}{8 M_{N}^2} |\psi_0(0)|^2$. Then, for $\psi_1$ at the origin, we have
\ali{
\label{eq:newaux}
\psi_1(0) = &-\frac{e^2 c_D}{8 M_N^2} |\psi_0(0)|^2 \int d^3x' G_0(0,\vec{x}';E) \psi_0(\vec{x}') \nonumber \\
& + \frac{e^2 c_D}{8 M_N^2} \psi_0(0) G_0(0,0;E).
}
The divergence appears in the last line. Indeed, it is well known that zero- and one-photon exchanges are power-like and logarithmically divergent, respectively \cite{Beneke:2013jia}. In dimensional regularization (with dimensions $d=4-2 \epsilon$), the latter leads to an ultraviolet pole $-Z^2 \alpha^2 m_{\mu}^2 r_n^2/3 \epsilon$ multiplying the LO decay rate,
where we used \cite{Beneke:2013jia}
\ali{
G_0(0,0;E) = \frac{Z \alpha m_{\mu}^2}{4 \pi \epsilon} + \mathcal{O}(\epsilon^0).
}

\textit{Hard function.}
That pole is canceled by a pole of IR origin in the hard function. To see its appearance, we focus on muon conversion and consider the 2-loop Feynman diagrams with photon exchanges between muon and nucleus, depicted in the first line of \ref{fig:diagrams} (the treatment in DIO is equivalent, with the relevant diagrams also shown in the figure).
We take the soft limit for the photons and the non-relativistic limit for the nucleus and muon. In that case, we use
\ali{
&\dfrac{1}{l_1^0 + i 0} \dfrac{1}{l_1^0 + l_2^0 + i 0}
+
\dfrac{1}{l_2^0 + i 0} \dfrac{1}{l_1^0 + l_2^0 + i 0} + \nonumber \\
&
\dfrac{1}{-l_1^0 + i 0} \dfrac{1}{-l_1^0 - l_2^0 + i 0}
+
\dfrac{1}{-l_2^0 + i 0} \dfrac{1}{-l_1^0 - l_2^0 + i 0} + \nonumber \\
&
\dfrac{1}{l_2^0 + i 0} 
\dfrac{1}{-l_1^0 + i 0} 
+
\dfrac{1}{l_1^0 + i 0} 
\dfrac{1}{-l_2^0 + i 0}
=
-4 \pi^2 \delta(l_1^0) \delta(l_2^0)
}
to write the muon conversion amplitude as
\ali{
\label{eq:2-loop-amplitude}
i \mathcal{M} = i C \frac{e^4 (2 m_{\mu})^2}{(2 \pi)^{2d}} 4 \pi^2 Y,
}
where $C$ represents the high-energy CLFV effective interaction (black disk in figure \ref{fig:diagrams}), and 
\ali{
Y \! \equiv \! \int \! \frac{d^dl_1 d^dl_2}{l_1^2 l_2^2} \frac{\mathcal{F}(l_1) \mathcal{F}(l_2) \delta(l_1^0) \delta(l_2^0)}{\left((p-l_1)^2-m_{\mu}^2\right)\left((p-l_1-l_2)^2 - m_{\mu}^2\right)} ,
}
with $\mathcal{F}$ being the FNS form factor (gray disk in figure \ref{fig:diagrams}). To determine the IR pole, it is enough to consider the low momentum expansion of the form factor,
\ali{
\label{eq:myFapprox}
\mathcal{F}(l) = -Z \left(1 + \frac{1}{6} r_n^2 l^2\right) + \mathcal{O}((l^2)^2),
}
focusing on the contributions proportional to $r_n^2$, and regulating the otherwise scaleless integrals with an upper bound. The $r_n^2$ term in $\mathcal{F}(l_1)$ leads to
\ali{
Y \big|_{r_n^2} = \frac{Z^2 \pi^4 r_n^2}{6 \epsilon} + \mathcal{O}(\epsilon^0).
}
Replacing in eq.~(\ref{eq:2-loop-amplitude}), we obtain
\ali{
i \mathcal{M}\big|_{r_n^2} = i C \frac{Z^2 \alpha^2 r_n^2 m_{\mu}^2}{6 \epsilon} + \mathcal{O}(\epsilon^0),
}
implying a pole $Z^2 \alpha^2 m_{\mu}^2 r_n^2/3 \epsilon$ multiplying the LO decay rate, precisely canceling the pole from the wavefunction.

\vspace{2mm}
\noindent
\textbf{Estimates.} 
Renormalizing $G_0(0,0,E)$ in $\overline{\text{MS}}$ and using the reduced Coulomb Green's function, we find
\ali{
\psi_1(0) &= \psi_0(0) \frac{2}{3}\left(Z \alpha m_{\mu} r_n\right)^2 \log \left(\frac{2 Z \alpha m_{\mu}}{\mu}\right).
}
Choosing the central scale $\mu = m_{\mu}$, the correction amounts to 
$-\left(5.34^{+2.23}_{-2.23}\right)\%$ relative to the LO decay rate, where the superscript 
and subscript indicate the variations obtained for $\mu = 2m_{\mu}$ and 
$\mu = m_{\mu}/2$, respectively.

Finally, we can assume a toy model for the form factor to ascertain the importance of finite size corrections to the hard functions.
We assume the model
\ali{
\label{eq:myFmodel}
\mathcal{F}(q) = -Z \left(1 - \dfrac{q^2 r_n^2}{12} \right)^{-2},
}
which is a simple expression compatible with eq.~(\ref{eq:myFapprox}).
At LO in muon DIO, and with $q^2 = - m_{\mu}^2$, we conclude that the FNS effects due to the contribution of $\mathcal{F}$ to the hard function lead to a $-55.38\%$ correction to the LO decay rate. Together with the effects due to $\psi_1$, this implies an overall correction of $-60.72\%$ to the muon DIO LO decay rate due to FNS effects.

\bibliographystyle{unsrt}
\bibliography{FS-letter}

\end{document}